\documentstyle[12pt]{article}
\begin{document}

\renewcommand{\baselinestretch}{1.5}
\newcommand\beq{\begin{equation}}
\newcommand\eeq{\end{equation}}
\newcommand\bea{\begin{eqnarray}}
\newcommand\eea{\end{eqnarray}}
\newcommand\phnx{\phi^x_n}
\newcommand\phny{\phi^y_n}
\newcommand\phnz{\phi^z_n}
\newcommand\psn{\psi_n}
\newcommand\sgnx{\sigma^x_n}
\newcommand\sgny{\sigma^y_n}
\newcommand\sgnz{\sigma^z_n}
\newcommand\vphn{{\vec \phi}_n}
\newcommand\vsgn{{\vec \sigma}_n}

\centerline{\bf A Majorana Fermion $t-J$ Model in One Dimension}
\vskip 1 true cm

\centerline{Diptiman Sen \footnote{E-mail address: diptiman@cts.iisc.ernet.in}}
\small 
\centerline{\it Centre for Theoretical Studies, Indian Institute of Science,}
\centerline{\it Bangalore 560012, India}
\vskip .25 true cm
\normalsize
\centerline{and}
\centerline{B. Sriram Shastry \footnote{E-mail address: 
bss@physics.iisc.ernet.in}}
\small
\centerline{\it Physics Department, Indian Institute of Science,}
\centerline{\it Bangalore 560012, India}
\normalsize
\vskip .5 true cm

\noindent 
{\bf Abstract}
\vskip .5 true cm
We study a rotation invariant Majorana fermion model in one dimension using 
diagrammatic perturbation theory and numerical diagonalization of small 
systems. The model is inspired by a Majorana representation of the 
antiferromagnetic spin-$1 \over 2$ chain, and it is similar in form to the 
$t-J$ model of electrons, except that the Majorana fermions carry spin-$1$ and 
$Z_2$ charge. We discuss the implications of our results for the low-energy 
excitations of the spin-$1 \over 2$ chain. We also discuss a generalization of 
our model from $3$ species of Majorana fermions to $N$ species; the $SO(4)$ 
symmetric model is particularly interesting.

\vskip .5 true cm
\noindent PACS numbers: ~71.10.Fd, ~75.10.Jm, ~75.50.Ee
 
\newpage

\centerline{\bf I. INTRODUCTION}
\vskip .5 true cm

In a recent paper \cite{BSS}, we used a representation of spin-$1 \over 2$ in 
terms of three species of Majorana fermions \cite{TSV,MAR} in order to study 
the antiferromagnetic spin-$1 \over 2$ chain. The Majorana represenation has an
advantage over other representations (such as the Schwinger boson or fermion
representations \cite{SCH,ABRI}) in that one does not have to impose a
constraint on the total particle number at each site (see however ref. 
\cite{MUD}). It is also rotation invariant unlike the "drone fermion" and 
the Holstein-Primakoff boson representations \cite{MATTIS,KENAN}.

For the spin-$1 \over 2$ chain with isotropic nearest neighbor interactions, 
the Majorana representation followed by a rotation invariant Hartree-Fock (H-F)
analysis \cite{BSS} leads to a picture of the low-energy excitations of the 
spin-$1 \over 2$ chain which is qualitatively similar to that obtained by
other methods \cite{dCP,FAD,MUL}. In particular, we find that the excitations 
are described by a two-parameter continuum in the $(q,\omega)$ space; for each 
momentum $q$, the low-energy spectrum has a range of energies $\omega$ as if 
the excitations are made up of two particles (called "spinons"). We also get 
reasonable dynamic structure functions and susceptibilites at all temperatures 
if we introduce some phenomenological structure functions. We should note
however that our Majorana fermions carry spin-$1$ unlike the ``standard''
spinons with spin-$1 \over 2$.

The positive features of the Majorana representation encourages us to study 
the fluctuations about the H-F state of the spin-$1 \over 2$ chain. More 
generally, it seems to be interesting to examine a strongly correlated 
Majorana fermion model in one dimension and contrast its properties with the
much better studied electronic systems like the Hubbard model. Such an
analysis would also be useful for other possible applications of Majorana 
fermions such as the Kondo problem \cite{TSV}. In this paper, we therefore 
study the $t-J$ model with Majorana fermions; the electronic version of this 
model has played a major role in theories of strongly correlated systems like
the high-temperature superconductors.

An outline of our paper is as follows. In section II, we briefly recall the 
Majorana representation of spin-$1 \over 2$ and the H-F analysis of the 
antiferromagnetic chain given in our earlier paper \cite{BSS}. This motivates 
a study of the $t-J$ 
model which is introduced in section III. We present the Feynman rules for 
the propagator and the vertex, and compute the one-loop correction to the
propagator. In section IV.A, we compute the two-loop correction to the 
propagator; we find the remarkable result that the on-shell correction
is of the same form as the tree level dispersion relation. In section IV.B,
we compute the two-loop correction to the dynamic structure function. The 
result can be used to perturbatively improve the power law of the equal-time 
correlation function and the ground state energy of the spin-$1 \over 2$ 
chain from the values obtained at the H-F level. In section IV.C, we study the 
one-loop correction to the vertex. In section V, we discuss the symmetries of 
the $t-J$ model and numerically analyze the spectrum of small systems using 
exact diagonalization. In section VI, we generalize our model from $SO(3)$ to 
$SO(N)$, and we briefly examine the $SO(4)$ case which is particularly 
interesting. Finally, in section VII, we summarize our understanding of the 
$t-J$ model.

\newpage
\centerline{\bf II. MAJORANA FERMIONS AND THE}
\centerline{\bf ANTIFERROMAGNETIC SPIN-$1 \over 2$ CHAIN}
\vskip .5 true cm

At each site $n$, the spin operators ${\vec S}_n = {\vec \sigma}_n / 2$ can
be written in terms of the Majorana operators $\vphn$ as \cite{BSS,TSV,MAR}
\bea
\sgnx ~&=&~ - ~i~ \phny ~\phnz ~, \quad \sgny ~=~ - ~i~ \phnz ~\phnx ~, 
\nonumber \\
{\rm and} \quad \sgnz ~&=&~ - ~i~ \phnx ~\phny ~.
\label{phi}
\eea
(We set Planck's constant equal to $1$). The hermitian operators $\phi^a_n$ 
(with $a=x,y,z$) satisfy the anticommutation relations
\beq
\{ ~{ \phi}^a_{m} ~,~ { \phi}^b_{n}~ \} ~=~ 2~ \delta_{mn} ~ \delta_{ab} ~.
\label{antip}
\eeq
Note that there is a local $Z_2$ gauge invariance since changing the sign of 
$\vphn$ does not affect ${\vec S}_n$. We will therefore say that $\vphn$ (or 
any odd power of it) carries a $Z_2$ charge. Let us define the trilinear and 
hermitian object $\psn = -i \phnx \phny \phnz$.
Then $[ \sigma^a_{m} , \psi_n ] = 0$, and $\{ \psi_m , \psi_n \} = 2 
\delta_{mn}$.  Under rotations, $\vphn$ and $\vsgn$ transform like vectors 
(spin-$1$ objects), while $\psn$ remains invariant. On the other hand, $\psn$ 
carries a $Z_2$ charge while $\vsgn$ is $Z_2$ neutral. Thus we have two 
different composite operators, $\vsgn$ and $\psn$, which carry spin and charge
respectively.

For a system with $L$ sites, it is known that the minimum possible dimension 
which allows a representation of the form given in equations 
(\ref{phi}-\ref{antip}) is $2^{L+[L/2]}$, where 
$[L/2]$ denotes the largest integer less than or equal to $L/2$. For $L$ 
sites with a spin-$1 \over 2$ object at each 
site, the Hilbert space clearly has a dimension $2^L$. Thus the Majorana 
representation of spin-$1 \over 2$ objects requires us to enlarge the space of 
states; the complete Hilbert space of states is given by a direct product of 
a `physical' space and an `unphysical' one. The operators $\vsgn$ act only on 
the physical states, while the $\vphn$ mix up different unphysical states. 

We now consider the Heisenberg antiferromagnetic chain with the Hamiltonian 
\beq
H ~=~ J ~\sum_n ~{\vec S}_n \cdot {\vec S}_{n+1} ~,
\label{ham}
\eeq
where $J > 0$. We use periodic boundary conditions ${\vec S}_{L+1} = {\vec 
S}_1$. The spectrum of (\ref{ham}) is exactly solvable by the Bethe ansatz; 
the ground state energy per site for large $L$ is given by
$E_0 /L= (- \ln 2 +1/4) J$ $= -0.4431 J$. The lowest excitations are known to
be four-fold degenerate consisting of a triplet ($S=1$) and a singlet ($S=0$) 
\cite{FAD}. The excitation spectrum is described by a two-parameter continuum 
in the $(q,\omega)$ space, where $- \pi < q \le \pi$. The lower boundary of the 
continuum is described by the des Cloiseaux-Pearson relation \cite{dCP}
\beq
\omega_l (q) ~=~ \frac{\pi J}{2} ~\vert ~\sin ~q ~\vert ~,
\label{oml}
\eeq
while the upper boundary is given by
\beq
\omega_u (q) ~=~ \pi J ~\vert ~\sin ~\frac{q}{2} ~\vert ~.
\label{omu}
\eeq
We can understand this continuum by thinking of these excitations as being made
up of two spin-$1 \over 2$ objects ("spinons") with the dispersion \cite{FAD}
\beq
\omega_s (q) ~=~ \frac{\pi J}{2} ~\sin ~q ~,
\label{disp1}
\eeq
where $0 < q < \pi$.
A triplet (or a singlet) excitation with momentum $q$ is made up of two 
spinons with momenta $q_1$ and $q_2$, such that $0 < q_1 \le q_2 < \pi$, 
$q=q_1 + q_2$, and $\omega (q) = \omega_s (q_1) +\omega_s (q_2)$. 

The Majorana analysis of this system proceeds as follows \cite{BSS}. We write
(\ref{ham}) in terms of Majorana operators and then perform a H-F 
decomposition. Thus
\bea
H & = & ~-~  \frac{J}{4} ~\sum_n ~(~ \phnx \phny \phi^x_{n+1} \phi^y_{n+1} ~+~
\mbox{cycl. perm. } (x,y,z) ~) \nonumber \\
& \simeq & ~\frac{J}{4} ~\sum_n ~[~ \phnx \phi^x_{n+1} \langle \phny 
\phi^y_{n+1} \rangle ~+~
\langle  \phnx \phi^x_{n+1} \rangle \phny  \phi^y_{n+1} ~-~ \nonumber \\
& & \quad \quad \quad \quad \langle \phnx \phi^x_{n+1} \rangle \langle \phny 
\phi^y_{n+1} \rangle ~+~ \mbox{cycl. perm. } (x,y,z)~] ~.
\label{hamhf1}
\eea
For a rotation and translation invariant H-F analysis, we have $g = i \langle 
\phi^a_{n} \phi^a_{n+1} \rangle$, where $g$ has the same value for all $n$ and 
$a=x,y,z$. (Our earlier paper \cite{BSS} follows slightly different 
conventions). The Fourier expansion of $\vphn$ is defined as
\beq
\phi^a_{n} ~=~ {\sqrt{2 \over L}} ~\sum_{0<q<\pi} ~[~ b_{aq} ~
e^{iqn} ~+~ b_{aq}^{\dag}~ e^{-iqn} ~] ~,
\label{fourier}
\eeq
where $\{b_{aq} , b_{bq^{\prime}}^{\dag} \} =\delta_{ab} \delta_{qq^\prime}$. 
We will work with {\it antiperiodic} boundary conditions for $\phi^a_{n}$ 
and {\it even} values of $L$ in order to eliminate modes with $q$ equal to 
$0$ and $\pi$. In equation (\ref{fourier}), $q=2\pi (p-1/2) 
/L$, with $p= 1,2,...,L/2$. In the limit $L \rightarrow \infty$, we get 
\beq
H ~=~  ~\sum_{a} ~\sum_{0<q<\pi} ~\omega_q ~b_{aq}^{\dag} b_{aq} ~+~ 
3LJ ~(~ \frac{g^2}{4}~ -~ \frac{g}{\pi} ~)~,
\label{hfham2}
\eeq
where the Majorana fermions have the dispersion $\omega_q = v \sin q$, with 
$v=2gJ$. The value of $g$ is determined self-consistently to be $g = 2/\pi$.
The H-F ground state energy is therefore 
\beq 
\frac{E_{0~HF}}{L} ~=~ - ~\frac{3}{\pi^2} ~J ~=~ - ~0.3040 ~J ~,
\label{e0}
\eeq
which is greater than the exact value mentioned above. The "spinon" spectrum 
has the same form as in (\ref{disp1}), except that we get $v = 4J/ \pi$
instead of $v_{exact}= \pi J /2$. 

We can go on to show that the Majorana fermion has spin-$1$, and a two-fermion 
state therefore has $S=0,1$ or $2$ in general. However the state created by 
$S^z_{q} = \sum_n ~S^z_{n} e^{-iqn}~$, where $0 < q < \pi$, has the form 
\beq
S^z_q ~\vert ~0~ \rangle ~=~ -i ~\sum_{\pi - q < k < \pi} ~b_{x,k}^{\dag} 
b_{y,q-k}^{\dag} ~\vert ~0~ \rangle ~,
\label{sqz}
\eeq
and has $S=1$. We thus obtain a two-parameter continuum of triplet excitations 
as in equations (\ref{oml}-\ref{omu}), with a prefactor $4/\pi$ instead of 
$\pi /2$.

Finally, the equal-time two-spin correlation function is given by
\bea
G_n ~\equiv ~ \langle ~0~ \vert ~{\vec S}_n \cdot {\vec S}_0 ~\vert ~0~ 
\rangle &=&~ \frac{3}{4} \quad {\rm for} \quad n=0 ~, \nonumber \\
&=&~ -~ \frac{3}{2\pi^2 n^2} ~[ 1 - (-1)^n ] \quad {\rm for} \quad n \ne 0~.
\label{corr}
\eea
This does not agree with the correct asymptotic behavior of $G_n$ which is
known to oscillate as $(-1)^n / n$. In particular, the H-F static structure 
function $S(q) = \sum_n G_n e^{-iqn}$ does not diverge as $q \rightarrow \pi$ 
in contrast to the correct $S(q)$ which has a logarithmic divergence at $\pi$.
(Note that we do get $\sum_n G_n =0$, as expected for a singlet ground state). 
We will show in section IV.B that two-loop effects effectively reduce the power
governing the asymptotic decay from $2$ to $1.75$ which is somewhat closer
to the correct value of $1$. At the same time, the ground state energy
per site is reduced from $-0.3040 J$ to $-0.3338 J$ which is also closer to
the Bethe ansatz value of $-0.4431 J$.

One can now consider fluctuations about the H-F ground state by doing loop 
calculations. However, instead of studying only the Hamiltonian (\ref{hamhf1}) 
as is sufficient for the spin-$1 \over 2$ chain, it is useful to study a more
general model which has the same structure but has two parameters instead
of one; the parameters are a hopping amplitude $t$ and a quartic
interaction $J$. This is the subject of the following sections.

\vskip .5 true cm
\centerline{\bf III. THE MAJORANA $t-J$ MODEL}
\vskip .5 true cm

We consider the Hamiltonian
\beq
H ~=~ \frac{-it}{4} ~\sum_{a,n} ~\phi^a_n \phi^a_{n+1} ~-~  \frac{J}{4} ~
\sum_n ~(~ \phnx \phny \phi^x_{n+1} \phi^y_{n+1} ~+~ \mbox{cycl. perm. }
(x,y,z) ~) ~,
\label{tj}
\eeq
with $t$ chosen to be positive, and we perform a perturbative expansion with 
the quartic term. To begin the diagrammatic analysis, we generalize the 
Fourier expression (\ref{fourier}) to the interaction picture field
\bea
\phi^a_n (t) ~&=&~ {\sqrt{2 \over L}} ~\sum_{- \pi < q < \pi} \phi^a_q ~
e^{i(qn - \omega_q t)} ~, \nonumber \\
{\rm where} \quad \phi^a_q ~&=&~ b_{aq} \quad {\rm if} \quad 0 < q < \pi ~, 
\nonumber \\
&=&~ b_{a,-q}^{\dag} \quad {\rm if} \quad - \pi < q < 0 ~, 
\label{heisen}
\eea
with 
\beq
\omega_q ~=~ t ~\sin q 
\label{disp3}
\eeq
for all $q$. Then we obtain the propagator 
\bea
\langle 0 \vert ~T ~\phi^a_q (t) ~\phi^b_{-q} (0) ~\vert 0 \rangle ~&\equiv&~ 
i ~ G^{ab} (q,t) ~=~ i~ \delta^{ab} ~G (q,t) ~, \nonumber \\
{\rm and} \quad i G (q, \omega) ~&=&~ i ~\int_{-\infty}^{\infty} ~dt ~
G(q,t) ~ e^{i \omega t} \nonumber \\
&=&~ \frac{i}{\omega - \omega_q + i \eta \theta (q)} ~,
\label{prop}
\eea
where $\eta$ is infinitesimal and positive, and $\theta (q) = 1$ if $0 < q 
< \pi$ and $-1$ if $-\pi < q < 0$.
For loop calculations, it is convenient to define a propagator even for
values of $q$ not lying in the range $[-\pi , \pi ]$. To do this,
we first define a momentum ${\underline q} = q + 2 n \pi$ where the integer $n$
is chosen such that $-\pi < {\underline q} \le \pi$. Then we {\it define}
$G(q,\omega) = G({\underline q},\omega)$ using (\ref{prop}). The propagator
is shown by a solid line in figure 1 (a).

The vertex shown in figure 1 (b) is obtained by Fourier transforming the 
quartic term in (\ref{tj}). The Feynman rule for the vertex is found to be
\bea
&& i \Gamma (a_1,q_1,\omega_1; a_2,q_2,\omega_2; a_3,q_3,\omega_3; a_4,q_4,
\omega_4) \nonumber \\
&& ~=~ i (2 \pi)^2 ~\delta_P (\sum_i q_i) ~\delta (\sum_i \omega_i) ~
4J ~\cos (\frac{1}{2} \sum_i q_i) ~\cdot \nonumber \\
&& \quad \quad ~\cdot ~\Bigl[ ~\delta^{a_1 a_2} ~ \delta^{a_3 a_4} ~\sin 
\frac{1}{2} (q_1 - q_2) ~\sin \frac{1}{2} (q_3 - q_4) 
\nonumber \\
&& \quad \quad \quad ~+~ \mbox{cycl. perm. } (a_2, q_2; a_3,q_3; a_4,q_4) ~
\Bigr] ~,
\label{vertex}
\eea
where the spin indices $a_1$ to $a_4$ can take the values $x,y,z$, and the 
momenta $q_1$ to $q_4$ need not lie in the range $[-\pi,\pi]$. The periodic
$\delta$-function is defined as
\beq
\delta_P (q) ~=~ \sum_{n=-\infty}^{\infty} ~\delta (q - 2n \pi ) ~.
\eeq
The expression in (\ref{vertex}) is antisymmetric under the exchange of any
two labels $(a_i,q_i,\omega_i)$ and $(a_j,q_j,\omega_j)$; it also vanishes if 
all the indices $a_i$ are equal.

We now compute the simplest loop effect, namely, the one-loop contribution to 
the propagator shown in figure 2 (a). It is called one-loop because there is 
one energy-momentum we have to integrate over. To this order in $J$, the 
self-energy is found to have the energy independent form 
\beq
\Sigma^{(1)} (q,\omega) ~=~ \frac{4J}{\pi} ~\sin q ~,
\label{prop1}
\eeq
where the superscript $(1)$ denotes the order of the loop. Thus the dispersion
relation changes from (\ref{disp3}) to
\beq
\omega_q ~=~ (t + \frac{4J}{\pi}) ~\sin q ~.
\label{disp4}
\eeq
We will therefore use the expression (\ref{disp4}) in the propagator
(\ref{prop}) for all the loop calculations below. Note that we can recover the 
antiferromagnetic spin-$1 \over 2$ chain by setting $t=0$ in (\ref{tj}); 
equation (\ref{disp4}) then gives us precisely the H-F dispersion discussed 
in section II.

\vskip .5 true cm
\centerline{\bf IV. LOOP CALCULATIONS}

\vskip .5 true cm
\centerline{\bf A. Two-Loop Contribution to Propagator}
\vskip .5 true cm

We will now compute the two-loop diagram shown in figure 2 (b). The two energy 
integrals can be easily done using the identities
\bea
\int_{-\infty}^{\infty} \frac{d\omega}{2\pi} ~\frac{1}{\omega - \alpha
+ i \eta} ~\frac{1}{\omega - \beta - i \eta} ~&=&~ \frac{i}{\beta - \alpha
+ i \eta} ~, \nonumber \\
\int_{-\infty}^{\infty} \frac{d\omega}{2\pi} ~\frac{1}{\omega - \alpha
+ i \eta} ~\frac{1}{\omega - \beta + i \eta} ~&=&~ 0 ~, 
\eea
if $\alpha$ and $\beta$ are real. 

We then obtain the following expression for the self-energy
\beq
\Sigma^{(2)} (q,\omega) ~=~ - \frac{4J^2}{\pi^2} ~\int_{-\pi}^{\pi} ~
\int_{-\pi}^{\pi} dl_1~ dl_2 ~\frac{\sin^2 [q+ \frac{1}{2} (l_1 +l_2)] ~
\sin^2 [\frac{1}{2} (l_1 -l_2)]}{\omega + \omega_{l_1} +\omega_{l_2} - 
\omega_{l_1 + l_2 +q} ~\pm i \eta }~,
\label{self}
\eeq
where we take the upper sign $(i \eta)$ in the denominator if $- \pi < l_1 , 
l_2 < 0$ and $0 < {\underline {l_1 + l_2 + q}} < \pi$, and we take the lower 
sign $(-i \eta)$ if $0 < l_1 , l_2 < \pi$ and $- \pi < {\underline {l_1 + l_2 
+ q}} < 0$. It is clear at this point that 
\beq
\Sigma^{(2)} (-q , -\omega ) ~=~ \Sigma^{(2)} (q, \omega ) ~;
\eeq
this property of the self-energy can be shown to be true to all orders in $J$.
Further, $\Sigma^{(2)} (\pi - q, \omega) = \Sigma^{(2)} (q, \omega )$. Now let 
us choose $0 < q < \pi$ and find the on-shell dispersion relation to order 
$J^2$, namely,
\beq
\omega ~=~ (t + \frac{4J}{\pi} ) ~\sin q ~+~ \Sigma^{(2)} (q, \omega) ~.
\label{disp5}
\eeq
To this order in $J$, we can set $\omega = (t + 4J/\pi) \sin q$ in the second 
term on the right hand side of (\ref{disp5}) or, equivalently, in the 
denominator of (\ref{self}). We then find that the denominator in (\ref{self}) 
never crosses zero in the given ranges of $l_1$ and $l_2$; thus we can drop 
the $\pm i \eta$ and the integrals are purely real. We then {\it numerically} 
find that (\ref{self}) has the remarkably simple form
\beq
\Sigma^{(2)} (q, (t+\frac{4J}{\pi}) \sin q) ~=~ - ~0.467 ~\frac{4 J^2}{\pi^2 
(t+ \frac{4J}{\pi})} ~\sin q ~.
\eeq
for all $q$ in the range $[0,\pi]$. Thus the dispersion relation to order $J^2$
is
\beq
\omega ~=~ \Bigl( ~t + \frac{4J}{\pi} - 0.189 ~\frac{J^2}{t + \frac{4J}{\pi}} ~
\Bigr) ~\sin q ~.
\label{disp6}
\eeq
We find it surprising that the form of the dispersion relation remains the 
same even at two-loops, and suspect that this may be true to all orders in $J$. 

\vskip .5 true cm
\centerline{\bf B. Two-Loop Contribution to Dynamic Structure Function}
\vskip .5 true cm

We will compute the two-spin correlation function 
\beq
S_{zz} (q,\omega ) ~\equiv ~\mbox{Fourier transform of } \langle 0 \vert 
S^z_n (t) S^z_0 (0) \vert 0 \rangle
\label{sqom}
\eeq
to two loops. To any order, we can show that this function remains invariant 
under $(q,\omega ) \rightarrow (-q, -\omega )$. We can obtain the static
structure function (equal-time correlation function) $S_{zz} (q)$ by 
integrating
\beq
S_{zz} (q) ~=~ \int_{-\infty}^{\infty} ~\frac{d\omega}{2\pi} ~S_{zz} (q,
\omega) ~e^{i \omega t} ~,
\label{sq}
\eeq
and taking the limit $t \rightarrow 0^+$. This is a function of $\vert q 
\vert$, so it is sufficient to compute it for $0 < q < \pi$.

The lowest order result for the correlation function is obtained from the 
one-loop diagram in figure 3 (a). After doing the energy integral, we obtain
\bea
S_{zz}^{(1)} (q,\omega) = \frac{i}{2\pi}~ \int_{-\pi}^{\pi} dl_1 &&
\frac{1}{\omega - \omega_{l_1} - \omega_{q-l_1} + i \eta} \quad {\rm if}
\quad 0 < l_1 , {\underline {q - l_1}} < \pi \nonumber \\
&& \frac{1}{\omega_{l_1} + \omega_{q-l_1} - \omega + i \eta} \quad {\rm if}
\quad - \pi < l_1 , {\underline {q - l_1}} < 0 ~. \nonumber \\
&&
\label{sqom1}
\eea
For $- \pi < q < \pi$, we then obtain
\beq
S_{zz}^{(1)} (q) ~=~ \frac{\vert q \vert}{2\pi} ~.
\label{sq1}
\eeq
The Fourier transform of this gives the spatial correlation function in
(\ref{corr}).

At two loops, we have to compute the diagram given in figure 3 (b). After
performing the two energy integrations, we arrive at the expression
\bea
S_{zz}^{(2)} (q,\omega ) = && \frac{iJ}{4\pi^2} ~ \int_{-\pi}^{\pi} 
\int_{-\pi}^{\pi} dl_1 ~dl_2 ~\sin \Bigl[ \frac{1}{2} (l_1 + l_2) \Bigr] ~\sin 
\Bigl[ q + \frac{1}{2} (l_1 +l_2) \Bigr] ~\cdot \nonumber \\
&& \cdot \frac{1}{(e_1 -i\eta) (e_2 - i \eta)} \quad {\rm if} \quad 0 < 
l_1 , l_2, - ({\underline {l_1 + q}}), - ({\underline {l_2 + q}}) < \pi 
\nonumber \\
&& \cdot \frac{1}{(e_1 +i\eta) (e_2 + i \eta)} \quad {\rm if} \quad  0 < -
l_1 , - l_2 , {\underline {l_1 + q}}, {\underline {l_2 + q}} < \pi 
\nonumber \\
&& \cdot \frac{-1}{(e_1 -i\eta) (e_2 + i \eta)} \quad {\rm if} \quad 0 < 
l_1 , - l_2 , - ({\underline {l_1 + q}}), {\underline {l_2 +q}} < \pi 
\nonumber \\
&& \cdot \frac{-1}{(e_1 +i\eta) (e_2 - i \eta)} \quad {\rm if} \quad 0 < -
l_1, l_2 , {\underline {l_1 +q}} , - ({\underline {l_2 + q}}) < \pi 
\nonumber \\
{\rm where} \quad e_1 ~=~ && \omega + \omega_{l_1} - \omega_{l_1 + q} ~, 
\nonumber \\
\quad e_2 ~=~ && \omega + \omega_{l_2} - \omega_{l_2 + q} ~.
\label{sqom2}
\eea
We then get, for $0 < q < \pi$,
\bea
S_{zz}^{(2)} (q) &=& -~ \frac{J}{2 \pi^2 (t + \frac{4 J}{\pi})} ~I(q) ~, 
\nonumber \\
I (q) &=& \int_0^q \int_0^q dl_1 ~dl_2 \frac{\cos \Bigl[ \frac{1}{2} (l_1 + 
l_2) \Bigr] ~\cos \Bigl[q - \frac{1}{2} (l_1 + l_2) \Bigr]}{\sin l_1 + \sin 
l_2 + \sin (q-l_1) + \sin (q-l_2)} ~.
\label{sq2}
\eea
We find analytically that $I(q)$ vanishes as $q \rightarrow 0$ and numerically
that 
\beq
\int_0^{\pi} ~dq~ I (q) ~=~ 0 ~.
\eeq
These are consistency checks following from the facts that the ground state
is a singlet and that the two-spin correlation at the same spatial point is 
equal to $3/4$; we already know that the one-loop correlation in equation 
(\ref{corr}) satisfies these checks. 

We now use equation (\ref{sq2}) to derive some interesting numbers relating
to the antiferromagnetic spin-$1 \over 2$ chain. First of all, 
we can show analytically that $I(q)$ is finite for all $q$, while
$I^{\prime} (q)$ diverges logarithmically at $q=\pi$ with coefficient
$1$, namely,
\bea
I^{\prime} (q) ~=~ \ln \vert \pi - q \vert ~+~ \quad \mbox{nondivergent 
terms } ~, \nonumber \\
I (q) ~=~ I ( \pi ) ~+~ (q - \pi ) ~\ln \vert \pi - q \vert \quad {\rm as}
\quad q \rightarrow \pi ~.
\eea
At long distances, the leading term in the spatial correlation function 
$G_n = 3 \langle 0 \vert S_n^z S_0^z \vert 0 \rangle$ takes the form 
\beq
\int_0^{\pi}  \frac{dq}{\pi} ~I (q) ~\cos (qn) ~=~ - ~\frac{(-1)^n}{\pi
n^2} ~\ln n ~+~ O(\frac{1}{n^2}) \quad {\rm as} \quad n \rightarrow 
\infty ~.
\label{intsq}
\eeq
After adding this to the one-loop result, we see that the long distance
correlation function has an oscillatory term going as
\beq
G_n ~=~ (-1)^n ~\frac{3}{2\pi^2 n^2} ~\Bigl[ 1 + \frac{J}{\pi (t+ 
\frac{4J}{\pi})} ~\ln n ~+~ \cdot \cdot \cdot ~\Bigl]
\eeq
where the dots indicate contributions from more than two loops.
If we now {\it assume} that these higher order terms come with the 
right numerical factors to turn the sum into an exponential series,
we see that the long distance correlation decays as $(-1)^n n^{-\alpha}$,
where the exponent $\alpha$ goes as
\beq
\alpha ~=~ 2 ~-~ \frac{J}{\pi (t + \frac{4J}{\pi})} 
\eeq
to order $J$. For the spin-$1 \over 2$ chain, we must
set $t=0$; this gives $\alpha = 1.75$ to this order. 

The second interesting number for the spin-$1 \over 2$ chain which we can 
derive from (\ref{sq2}) is the ground state energy per site; this is equal to 
$JG_1$ for $t=0$. On numerically integrating (\ref{sq2}), we find the two-loop 
result
\beq
G_1^{(2)} ~=~ - ~\frac{3}{8 \pi^2} ~\int_0^{\pi} ~dq ~I (q) ~\cos q ~=~ - 
0.0298 ~.
\eeq
On adding this to the one-loop result, we get the value $-0.3338 J$.

\vskip .5 true cm
\centerline{\bf C. One-Loop Contribution to Vertex}
\vskip .5 true cm

For completeness, we will mention the one-loop correction to the vertex.
Let us choose two of the spin indices to be $x$ and two to be $y$. From 
(\ref{vertex}), the zero-loop form of the vertex is given by $(2\pi)^2$ 
times the energy-momentum conserving $\delta$-functions times
\beq
i \Gamma^{(0)} ~=~ i 4 J ~\cos (\frac{1}{2} \sum_i q_i) ~\sin \frac{1}{2} 
(q_1 - q_2) ~ \sin \frac{1}{2} (q_3 - q_4) ~.
\eeq
The one-loop correction
$i \Gamma^{(1)} (x,q_1,\omega_1; x,q_2,\omega_2; y,q_3,\omega_3; y,q_4,
\omega_4)$ is given by the sum of the three diagrams shown in figure 4. On
doing the energy integral, we find that the contribution of figure 4 (a) is
\bea
&& -i 8J^2 ~\cos (\frac{1}{2} \sum_i q_i) ~\sin \frac{1}{2} (q_1 - q_2) ~
\sin \frac{1}{2} (q_3 - q_4) ~\cdot \nonumber \\
&& \cdot ~\int_{-\pi}^{\pi} ~\frac{dl}{2\pi} ~\sin^2 [l + 
\frac{1}{2} (q_1 + q_2) ] ~\cdot \nonumber \\
&& \cdot ~\frac{1}{\omega_{l+q_1 + q_2} - \omega_l - \omega_1 - \omega_2 + i 
\eta} \quad {\rm if} \quad 0 < l, - ({\underline {l + q_1 + q_2}}) < \pi 
\nonumber \\
&& \cdot ~\frac{1}{\omega_l + \omega_1 + \omega_2 - \omega_{l+q_1 + q_2} + i 
\eta} \quad {\rm if} \quad 0 < - l, {\underline {l + q_1 + q_2}} < \pi ~.
\label{ver1a}
\eea
The contribution of figure 4 (b) can be obtained from equation (\ref{ver1a}) by
changing the coefficient $8$ to $16$ and cyclically replacing $q_2 
\rightarrow q_3 \rightarrow q_4 \rightarrow q_2$ and $\omega_2 \rightarrow 
\omega_3$. The contribution of figure 4 (c) can be obtained from equation 
(\ref{ver1a}) by changing $8$ to $16$ and replacing $q_2 \rightarrow q_4 
\rightarrow q_3 \rightarrow q_2$ and $\omega_2 \rightarrow \omega_4$. 

\vskip .5 true cm
\centerline{\bf V. SYMMETRIES AND NUMERICAL RESULTS}

\vskip .5 true cm
\centerline{\bf A. Numerical Results}
\vskip .5 true cm

We can numerically study the spectrum of our model by exact diagonalization of 
small systems. To do that, it is useful to know all the symmetries of the 
model. Some of the conserved quantum numbers are the total spin 
${\vec S}^2$ and any one of its components, say, $S_z$, the total momentum 
$q$ modulo $2\pi$, and parity $P = \pm 1$ which arises from the symmetry of 
the Hamiltonian under 
\beq
\vphn ~\rightarrow ~(-1)^n ~{\vec \phi}_{L+1-n} ~.
\eeq
In addition, there is a $Z_2$ quantum number defined as follows. Consider
\bea
\Gamma ~&=&~ \psi_1 \psi_2 \cdot \cdot \cdot \psi_L \quad {\rm if} \quad
\frac{L}{2} \quad \mbox{is even } \nonumber \\
&=&~ i ~\psi_1 \psi_2 \cdot \cdot \cdot \psi_L \quad {\rm if} \quad
\frac{L}{2} \quad \mbox{is odd } ~,
\label{gam}
\eea
satisfying $\Gamma^{\dag} = \Gamma^{-1} = \Gamma$. This operator anticommutes
with each of the $\vphn$ and therefore commutes with the Hamiltonian 
(\ref{tj}). Hence the eigenvalue of $\Gamma = \pm 1$ is a good quantum number.
We will {\it define} $\Gamma =1$ for the ground state of the $t-J$ model; we 
can ensure this by introducing a $-$ sign in the definition (\ref{gam}) if
necessary. 

There are a few selection rules and energy degeneracies connecting some
of these quantum numbers. We will see below that the ground state has
$q=0$, and we choose $\Gamma =1$. We can now obtain various excited states 
by acting on it with a certain number of Majorana operators as defined in 
(\ref{fourier}). Each such operator carries a momentum $q$ which is an
{\it odd} multiple of $\pi /L$, and $\Gamma =-1$. It is therefore clear 
that all states must have $\Gamma = \exp (iqL)$; this eiegenvalue is $1$ or
$-1$ depending on whether the state has an even or odd number of Majorana
fermions. Secondly, Majorana operators with momenta $q$ and $\pi -q$ carry 
the same energy by parity. Thus states with an odd number of Majoranas,
i.e. with $\Gamma =-1$, must have an energy degeneracy between total momenta
equal to $q$ and $\pi -q$. States with $\Gamma =1$ must have an energy 
degeneracy between momenta $q$ and $-q$.

For a numerical study, it is more convenient to rewrite (\ref{tj}) in the form
\beq
H ~=~ \frac{1}{4} ~\sum_{n=1}^L ~( ~-~ it ~\psn \psi_{n+1} ~+~ J ~) ~\vsgn 
\cdot {\vec \sigma}_{n+1} ~.
\label{tj2}
\eeq
As mentioned in section II, we use periodic boundary conditions for $\vsgn$ and 
antiperiodic for $\psn$. We diagonalize $H$ in a basis consisting of a direct 
product of states of the form $\vert \Psi_i \rangle \otimes \vert \alpha_j 
\rangle$, such that the operators $\vsgn$ and $\psn$ act only on $\vert \Psi_i
\rangle$ and $\vert \alpha_j \rangle$ respectively. In order to study the
spectral flow from the pure-$J$ model to the pure-$t$ model, we introduce a
parameter $x$ lying between $0$ and $1$, such that $J=4(1-x)$ and $t=4x$. Thus
\beq
H (x) ~=~ \sum_{n=1}^N ~( ~1 ~-~ x ~-~ i x ~\psn \psi_{n+1} ~)~ \vsgn \cdot
{\vec \sigma}_{n+1} ~.
\label{hx}
\eeq
We have obtained the eigenvalues of (\ref{hx}) for $L=4$ and $6$, for $11$
equally spaced values of $x$ from $0$ to $1$. All the conserved quantities 
discussed above have discrete eigenvalues; hence these remain invariant as
$x$ changes. Numerically, we only kept track of the eigenvalues of total spin 
$S=0, 1, ..., L/2$ and total $S_z =0$; whenever necessary, the eigenvalues 
$q$, $P$ and $\Gamma$ can be deduced by continuity arguments from the exact 
analytical solutions known at $x=1$. The energy eigenvalues in each 
$S$ sector are shown in figures 5 (a-c) for $L=4$, and the lowest few 
eigenvalues in each $S$ sector are shown in figures 6 (a-d) for $L=6$.
We should remark here that the degeneracies of the various levels have not 
been shown, and that we have not distinguished between true crossings and
avoided crossings in these figures.

To get a feeling for the elementary excitations, let us discuss six 
low-lying states marked $a-f$ on the figures; these include the three lowest 
states $a,b,c$ with $S=0$ in figures 5 (a) and 6 (a), the two lowest states 
$d,e$ with $S=1$ in figures 5 (b) and 6 (b), and the lowest state $f$ with 
$S=2$ in figures 5 (c) and 6 (c). The energy dependence of these six states 
can be seen to be quite similar for $L=4$ and $6$. The ground state, 
marked $a$, is unique for all values of $x$ (except $x=0$ where it has a 
degeneracy of $2^{L/2}$); it has spin $S=0$, momentum $q=0$, and $\Gamma =1$. 
The next two states in the $S=0$ sector, marked $b$ and $c$, have 
$\Gamma =1$ and $-1$ with degeneracies of $1$ and $2$ respectively; 
these two states exhibit a true level crossing between $x=0$ and $1$, so that 
$b$ is lower than $c$ near $x=1$ and vice versa near $x=0$.
The two states with $S=1$, marked $d$ and $e$, have $\Gamma =-1$ and $1$
with degeneracies of $2$ and $1$ respectively. These also exhibit a true level
crossing, with $d$ being lower than $e$ near $x=1$ and vice versa near $x=0$.
Finally, the state with $S=2$ marked $f$ has $\Gamma =1$ and is nondegenerate.

The composition of these six states can be easily understood at the 
noninteracting point $x=1$. At this point, the ordering of energies is given 
by $a < d < b=e=f < c$. The ground state $a$ is the empty state. State $d$
contains a single Majorana fermion with spin $1$, with momentum equal to 
either $\pi /L$ or $\pi - \pi /L$; hence the double degeneracy. The state $b$ 
contains two fermions in a spin-$0$ combination, one with 
momentum $\pi /L$ and the other with momentum $\pi - \pi /L$; hence the 
total momentum is $\pi$. States $e$ and $f$ have the same composition as $b$,
except that they have spins $1$ and $2$ respectively. State $c$ has three
fermions in a spin-$0$ combination, two with momenta $\pi/L$ and 
$\pi - \pi /L$, and the third with momentum either $3 \pi /L$ or $\pi - 3 
\pi /L$; the double degeneracy is due to the two-fold choice for the third 
fermion. If we now move from $x=1$ to $x=0$, all these states get "dressed"
with an even number of fermions. At $x=0$, the energy ordering is $a=b=c <
d=e < f$.

Although the system sizes are not large, we can draw the following qualitative 
conclusions from these figures.
First, the states evolve smoothly from $x=0$ to $x=1$ with no abrupt changes
in between. In each spin sector, the lowest energy states at $x=0$ are mainly
composed of the lowest energy states at $x=1$, and vice versa. Finally, the
complex pattern of level crossings for small values of $S$ seems to suggest
that the model is nonintegrable for $x$ not equal to $0$ or $1$.

\vskip .5 true cm
\centerline{\bf B. Conformal Field Theory: A Conjecture}
\vskip .5 true cm

It would be useful to understand the low-energy excitations of the model
in terms of conformal field theory; amongst other things, this would lead
to a simpler derivation of various correlation functions (see ref. \cite{AFF} 
and references therein). We would like to advance a conjecture in this 
direction. Before doing that, we must consider the two limits of the 
Hamiltonian (\ref{hx}) which are exactly solvable.

For $x=1$, we have three uncoupled Majorana 
fermions with the same dispersion (\ref{disp3}). The low-energy excitations 
(modes with momenta $q$ close to $0$ or $\pi$) have velocity $t=4$ and are 
governed by a conformal field theory which is an $SU(2)_2$ Wess-Zumino-Witten 
(WZW) model with central charge $c=3/2$. 

For $x=0$, the unphysical states decouple completely. The physical
states (each of which have an unphysical degeneracy of $2^{[L/2]}$ due to
the spinless Majorana field $\psn$) are solvable by the Bethe ansatz; the 
low-energy physical excitations have the velocity $\pi J /2 =2\pi$ and are 
governed by a $SU(2)_1$ WZW conformal 
field theory with $c=1$. The $x=0$ limit is somewhat singular
due to the complete decoupling of the unphysical states. Let us therefore
examine what happens if $x$ is nonzero but small. We can then do degenerate 
perturbation theory to first order in $x$. For instance, consider 
perturbation theory amongst the $2^{[L/2]}$ ground states which are 
degenerate for $x=0$; we denote these
states by the direct product $\vert \Psi_0 \rangle \otimes \vert \alpha 
\rangle$, where $\Psi_0$ is the physical ground state and $\alpha$ can take 
$2^{[L/2]}$ values. By rewriting $\vphn = \vsgn \psn$ and using the Bethe 
ansatz value
\beq
e ~\equiv~ \langle \Psi_0 \vert \vsgn \cdot {\vec \sigma}_{n+1} 
\vert \Psi_0 \rangle ~=~ - 1.7726 ~,
\eeq
we find that the first term in the Hamiltonian (\ref{tj}) can be written as
the perturbation
\beq
V ~=~ -ixe ~\sum_n ~\psn \psi_{n+1} ~.
\eeq
This can be diagonalized by Fourier transforming as 
\beq
\psn ~=~ {\sqrt {2 \over L}} ~\sum_{0<q<\pi} ~[ c_q^{\dag} e^{iqn} ~+~ c_q
e^{-iqn} ] ~.
\eeq
Then
\beq
V ~=~ -4xe ~\sum_{0<q<\pi} ~\sin q ~c_q^{\dag} c_q ~+~ \frac{2Lxe}{\pi} ~.
\eeq
Thus the spinless sector with $Z_2$ charge has low-energy excitations with
velocity $-4xe$. These are described by a conformal field theory with 
$c=1/2$. Thus the spin and charge excitations have completely different
velocities.

The question now is what happens in between the two limits. Although our 
numerical results are limited to $L=4$ and $6$, they suggest that both the 
spin sector (for instance, states with $S > 0$ and $\Gamma =1$) and the
charge sector (states with $S=0$ and $\Gamma =-1$) remain gapless for all
values of $x$; there does not appear to be a quantum phase transition at any
point between $x=0$ and $1$. It is then natural to conjecture that the 
low-energy sector is generally described by the product of two conformal 
field theories which have different velocities; the spin sector by a $SU(2)_1$ 
WZW model with $c=1$, and the $Z_2$ charge sector by a single Majorana fermion 
with $c=1/2$. If this is correct, it would be somewhat reminiscent of the 
one-dimensional 
Hubbard model away from half-filling; the low-energy excitations of this are 
governed by the product of two conformal field theories which have different 
velocities if the on-site interaction $U \ne 0$; the spin sector is again 
described by a $SU(2)_1$ WZW model while the $U(1)$ charge sector is described 
by a Gaussian field theory with $c=1$ \cite{AFF,WOY}.

\vskip .5 true cm
\centerline{\bf VI. $SO(N)$ $t-J$ MODEL}
\vskip .5 true cm

It is possible to generalize the $t-J$ model with three species of Majorana
fermions to a model with $N$ species. In terms of an interpolating parameter
$x$, we can write a $SO(N)$ symmetric Hamiltonian in the form 
\beq
H ~=~ -~ix ~\sum_{n} ~\sum_{a=1}^N ~\phi^a_n \phi^a_{n+1} ~-~ (1-x)~ \sum_n ~
\sum_{1\le a < b \le N} ~\phi^a_n ~\phi^b_n ~ \phi^a_{n+1} ~\phi^b_{n+1} ~, 
\label{tj3}
\eeq
where the operators $\phi^a_n$ satisfy the same anticommutation relations 
as in (\ref{antip}), except that the flavor indices $a,b$ can now take $N$ 
values. The Hilbert space for $L$ sites has the dimensionality $2^{NL/2}$ if 
$L$ is even. For $x=1$, we have $N$ noninteracting Majorana fermions with
the dispersion $\omega_q = 4 \sin q$; the low-energy excitations are therefore
described by a $c=N/2$ conformal field theory. We will now examine two special 
cases, $N=2$ and $N=4$, for which the antiferromagnetic limit $x=0$ is also 
well understood.

For $N=2$, the model is equivalent to the $XXZ$ spin-$1 \over 2$ chain. This 
can be shown as follows. We first combine two Majorana operators to produce an 
annihilation operator for a spinless Dirac fermion.
\beq
d_n ~=~ \frac{(-i)^n}{2} ~( ~\phi^1_n ~+~ i ~\phi^2_n ~) ~.
\eeq
These satisfy the anticommutation relation
\beq
\{ ~d_m ~,~ d_n^{\dag} ~\} ~=~ \delta_{mn} ~.
\eeq
In terms of these, the Hamiltonian takes the form 
\beq
H ~=~ 2x ~\sum_n ~\Bigl( ~d_n^{\dag} d_{n+1} ~+~ d_{n+1}^{\dag} d_n ~\Bigr) 
+~ 4 (1-x)~ \sum_n ~(~ d_n^{\dag} d_n ~-~ \frac{1}{2} ~)~(~ d_{n+1}^{\dag} 
d_{n+1} ~-~ \frac{1}{2} ~) ~.
\eeq
A Jordan-Wigner transformation from fermions to spin-$1 \over 2$ operators
then produces the $XXZ$ Hamiltonian \cite{AFF}
\beq
H ~=~ x ~\sum_n ~\Bigl( ~\sigma_n^x \sigma_{n+1}^x ~+~ \sigma_n^y
\sigma_{n+1}^y ~\Bigr) ~+~ (1-x) ~\sum_n ~\sigma_n^z \sigma_{n+1}^z ~.
\eeq
This model is exactly solvable by the Bethe ansatz for all values of $x$; it 
has a quantum phase transition at $x = 1/2$. For $1/2 \le x \le 1$, the model 
is gapless and is described by a $c=1$ Gaussian conformal field theory (the
symmetry is enhanced from $U(1)$ to $SU(2)$ at $x=1/2$). For $0 \le x < 1/2$, 
the model is gapped and has a Neel ground state with long range order.

The case $N=4$ is more interesting. At $x=0$, the model is a direct
sum of two antiferromagnetic spin-$1 \over 2$ chains. To show this, let
us first define the six generators of $SO(4)$ at each site,
\beq
K_n^{ab} ~=~ \frac{i}{2} ~\phi_n^a \phi_n^b ~.
\eeq
Now we use the homomorphism $SO(4) \simeq SO(3) \times SO(3)$. This can 
be proved by defining the linear combinations 
\bea
L_{1n}^x ~&=&~ \frac{1}{2} ~(~ K_n^{23} ~+~ K_n^{14} ~)~, \quad \quad
L_{2n}^x ~=~ \frac{1}{2} ~(~ K_n^{23} ~-~ K_n^{14} ~)~, \nonumber \\
L_{1n}^y ~&=&~ \frac{1}{2} ~(~ K_n^{13} ~-~ K_n^{24} ~)~, \quad \quad
L_{2n}^y ~=~ \frac{1}{2} ~(~ K_n^{13} ~+~ K_n^{24} ~)~, \nonumber \\
L_{1n}^z ~&=&~ \frac{1}{2} ~(~ K_n^{12} ~+~ K_n^{34} ~)~, \quad \quad
L_{2n}^z ~=~ \frac{1}{2} ~(~ K_n^{12} ~-~ K_n^{34} ~)~. 
\label{l12}
\eea
These generate two commuting $SO(3)$ algebras, namely,
\beq
[~ L_{\alpha m}^a ~,~ L_{\beta n}^b ~] ~=~ i ~\delta_{\alpha \beta} 
\delta_{mn} ~\sum_c ~\epsilon^{abc} ~L_{\alpha m}^c ~,
\eeq
where $\alpha , \beta =1,2$ label the two algebras, $a,b,c =x,y,z$, and
$\epsilon^{xyz} =1$. We can define total angular momentum operators
\beq
L_{\alpha}^a ~=~ \sum_n ~L_{\alpha n}^a ~;
\eeq
these commute with the Hamiltonian (\ref{tj3}) for all values of $x$.

At a single site, the Hilbert space is four-dimensional; the four operators 
$\phi^a$ can be chosen to be the $\gamma$ matrices used in Dirac's theory of 
the electron. One can verify that
\bea
{\vec L}_1^2 ~=~ \frac{3}{8} ~(~ I ~-~ \phi_1 \phi_2 \phi_3 \phi_4 ~)~, 
\nonumber \\
{\vec L}_2^2 ~=~ \frac{3}{8} ~(~ I ~+~ \phi_1 \phi_2 \phi_3 \phi_4 ~)~.
\eea
It is convenient to choose a representation in which these two operators
are diagonal in the form of $2 \times 2$ blocks
\bea
{\vec L}_1^2 ~&=&~ \left( {\begin{array}{cc}
                          3/4 & 0 \\
                          0 & 0
                          \end{array}} \right) ~, \nonumber \\
{\vec L}_2^2 ~&=&~ \left( {\begin{array}{cc}
                      	  0 & 0 \\
                          0 & 3/4
                          \end{array}} \right) ~.
\eea
Thus the upper two components of the Hilbert space transform as the
$(\frac{1}{2},0)$ representation of $({\vec L}_1,{\vec L}_2)$, while the 
lower two components transform as $(0,\frac{1}{2})$. We now see that, for 
$x=0$, the Hamiltonian for $L$ sites has the block diagonal form
\beq
H ~=~ \left( {\begin{array}{cc}
             H_1 & 0 \\
             0 & H_2
             \end{array}} \right) ~, \nonumber 
\label{hbd}
\eeq
where the Hamiltonians $H_1$ and $H_2$ act on two separate $2^L$ dimensional 
Hilbert spaces, each corresponding to a spin-$1 \over 2$ chain. Here
\beq
H_{\alpha} ~=~ 2J ~\sum_n ~{\vec L}_{\alpha,n} \cdot
{\vec L}_{\alpha,n+1} ~,
\label{hal}
\eeq
for $\alpha =1,2$. We already know that this can be solved by the Bethe
ansatz; the block diagonal form of (\ref{hbd}) implies that each eigenvalue 
will have a two-fold degeneracy.  Thus the $SO(4)$ $t-J$ model is exactly
solvable at both $x=0$ and $1$, and one can investigate how the spectrum 
interpolates between the two. We will not pursue this here.

The Majorana fermions in the $SO(4)$ model carry the spin quantum numbers
$(L_1 , L_2)=(\frac{1}{2},\frac{1}{2})$. In this respect they may be closer 
in spirit to the Faddeev-Takhtajan spinons (which are spin-$1 \over 2$ 
objects) than the Majorana fermions in the $SO(3)$ model which carry spin-$1$. 
To show this more precisely, let us define two Dirac fermion operators in the 
$SO(4)$ model as
\bea
d_{1n} ~&=&~ \frac{(-i)^n}{2} ~(~\phi_{1n} ~+~i \phi_{2n} ~)~, \nonumber \\
d_{2n} ~&=&~ \frac{(-i)^n}{2} ~(~\phi_{3n} ~+~i \phi_{4n} ~)~.
\label{dir}
\eea
We can then verify that the particles created by $d_{1n}^{\dag}$ and
$d_{2n}^{\dag}$ have the eigenvalues of the total angular momentum operators
$(L_1^z , L_2^z)$ equal to $(\frac{1}{2},\frac{1}{2})$ and $(\frac{1}{2}, -
\frac{1}{2})$ respectively. Thus, for the purely antiferromagnetic model 
$x=0$, a fermion operator acting on the ground state of, say, the $L_1$ chain 
will produce states which transform 
as spin-$1 \over 2$ under the operators ${\vec L}_1$; in addition, the states
will carry a two-fold internal quantum number coming from ${\vec L}_2$.

It is interesting to note that the hopping term (proportional to $t$) in the 
$SO(4)$ Majorana model is identical to the hopping term in the Hubbard model 
of electrons. However the four-fermion interactions are very different in the 
two models.

Before ending this section, we would like to mention that a H-F analysis
of the $SO(N)$ antiferromagnet has been performed in ref. \cite{FOE}. Their
H-F decomposition differs from the one we have used in section II. Consequently
they obtain a much higher value for the ground state energy than us, namely, 
equation (\ref{e0}) for $N=3$, and $-JN(N-1) /2\pi^2$ in general.

\vskip .5 true cm
\centerline{\bf VII. DISCUSSION}
\vskip .5 true cm

We have studied a one-dimensional $SO(3)$ invariant $t-J$ model with Majorana 
fermions. At the pure-$J$ end, this describes the nearest neighbor 
antiferromagnetic spin-$1 \over 2$ chain, while at the pure-$t$ end, we have 
three noninteracting fermions. We have done perturbative calculations to low 
order in the four-fermion interaction. We have also studied the model 
numerically by exact diagonalization of small systems. These studies provide a 
new perspective on the excitations of the spin-$1 \over 2$ chain by relating 
it in an "adiabatic" and rotationally invariant way to a model of free 
Majorana fermions. The low-energy excitations of the spin-$1 \over 2$ chain 
can be thought of as being made up of a small number of Majorana fermions.

The field theoretic description of the low-energy excitations of the 
model remains unclear. We have suggested that these excitations are governed 
by a product of two conformal field theories which have entirely different 
symmetries. Numerical studies, particularly finite size scaling, of much 
larger systems are required to test this scenario.

The $SO(N)$ generalization of our model also deserves further study. The 
$SO(4)$ case seems to be specially interesting because it provides yet 
another way of smoothly connecting a model of free fermions to a spin-$1 
\over 2$ antiferromagnet. The $SO(4)$ model may also have applications to the
problem of two coupled spin-$1 \over 2$ chains \cite{SHE}.

\vskip .5 true cm
\centerline{\bf ACKNOWLEDGMENTS}
\vskip .5 true cm

DS thanks the Department of Physics and Astronomy, McMaster University for 
its hospitality during part of this work.

\vskip .5 true cm

\newpage

\noindent {\bf Figure Captions}
\vskip 1 true cm

\noindent 
{1.} The propagator and vertex for the Majorana $t-J$ model.

\noindent
{2.} The one- and two-loop contributions to the propagator.

\noindent
{3.} The one- and two-loop contributions to the two-spin correlation function.

\noindent
{4.} The one-loop contributions to the vertex.

\noindent
{5.} Energies for $L=4$. $x=0$ and $1$ denote the pure-$J$ and pure-$t$ models 
respectively. Figures (a-c) show all the energies for total $S=0,1,2$. The 
curves marked $a-f$ are discussed in the text.

\noindent
{6.} Energy for $L=6$. Figures (a-d) show the lowest $20$ energies for $S=0,1,
2$ and all the energies for $S=3$. The curves marked $a-f$ are discussed in 
the text.

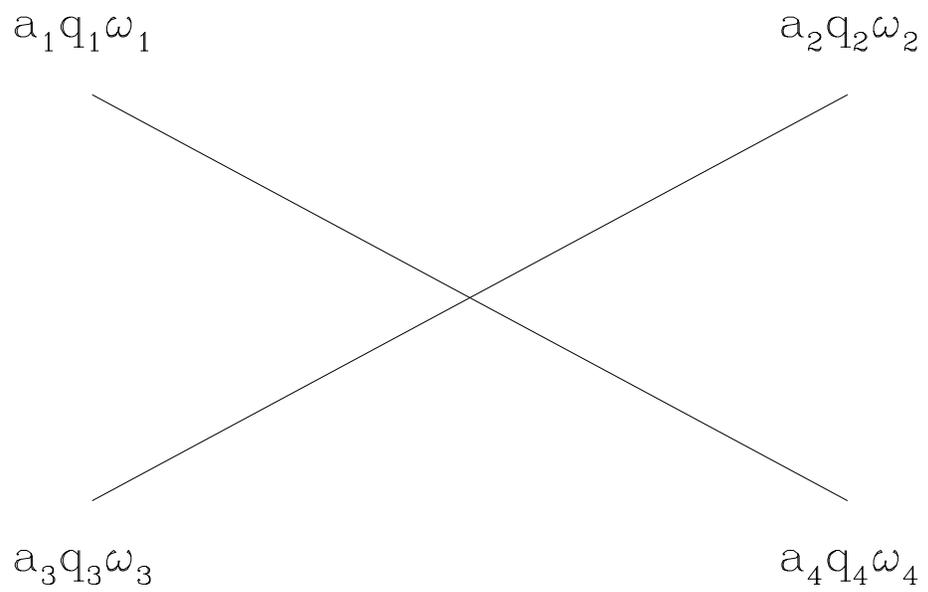
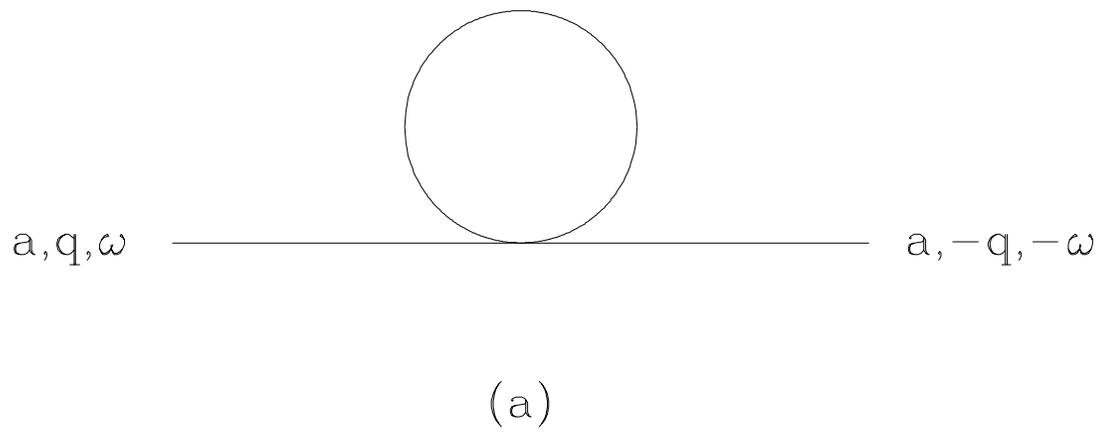
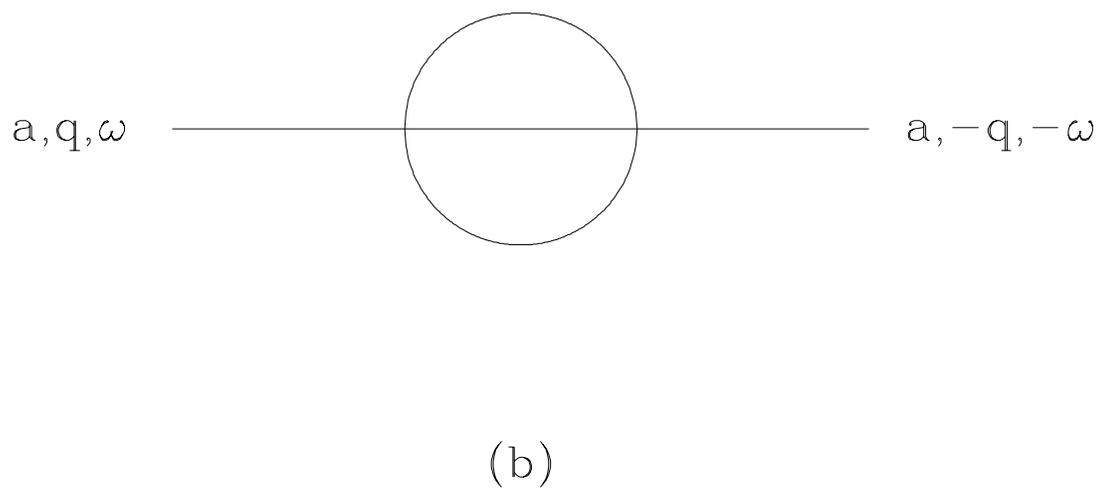
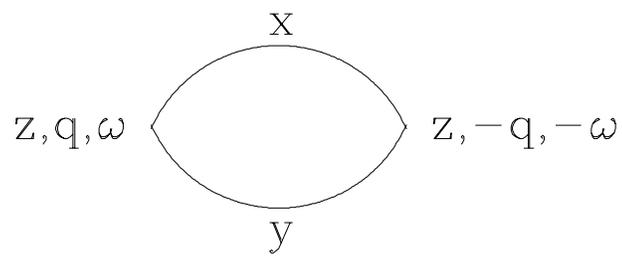
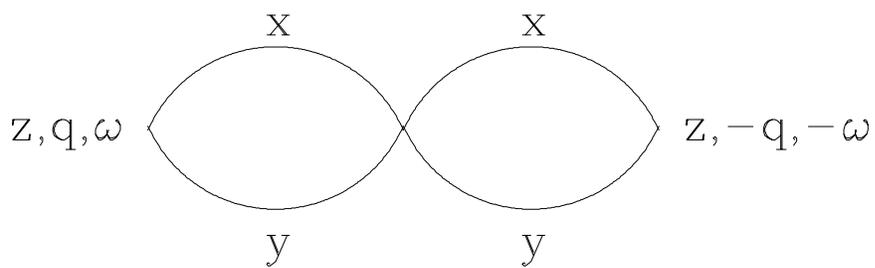
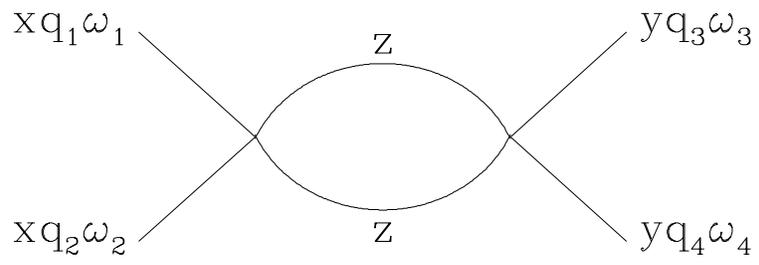
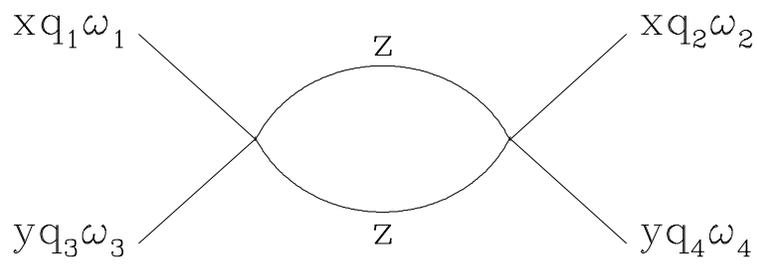
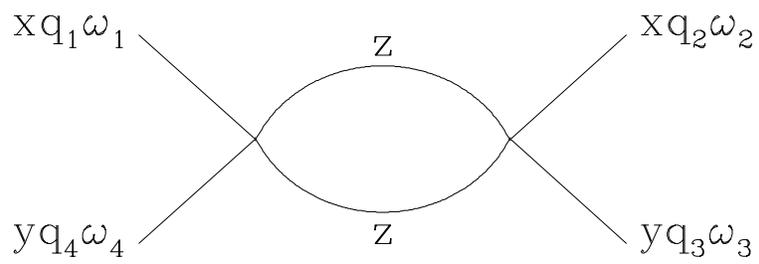
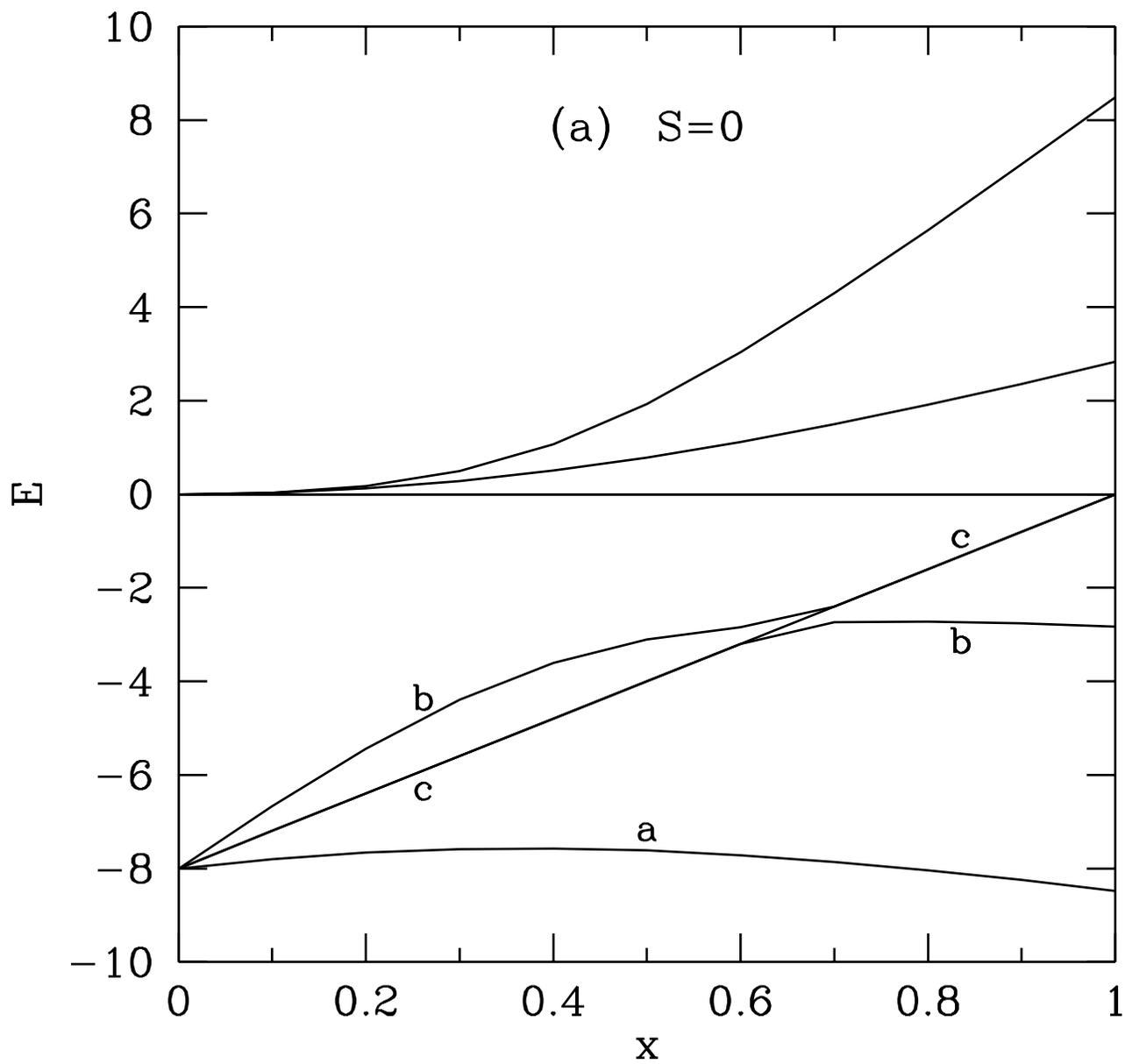
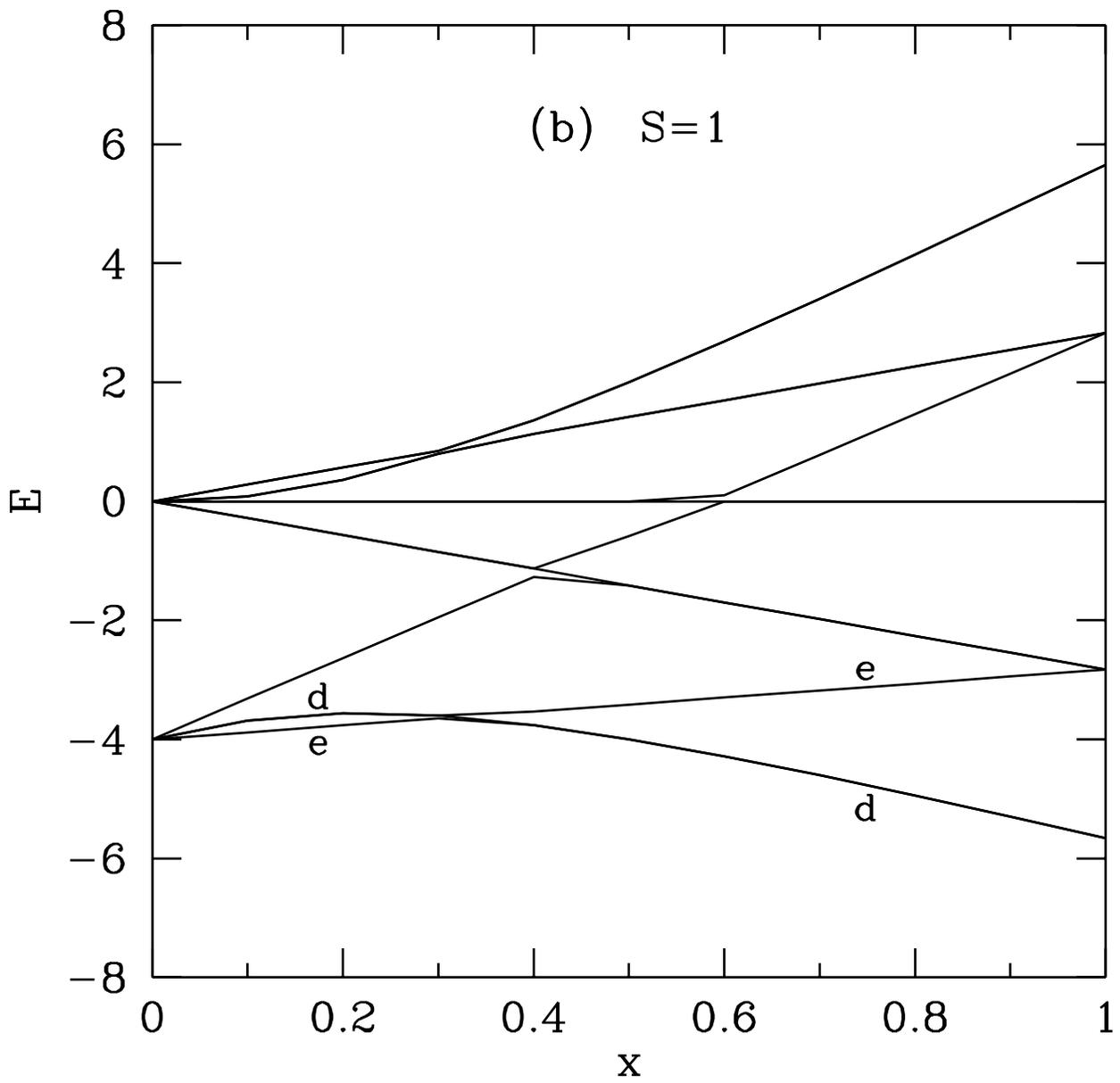
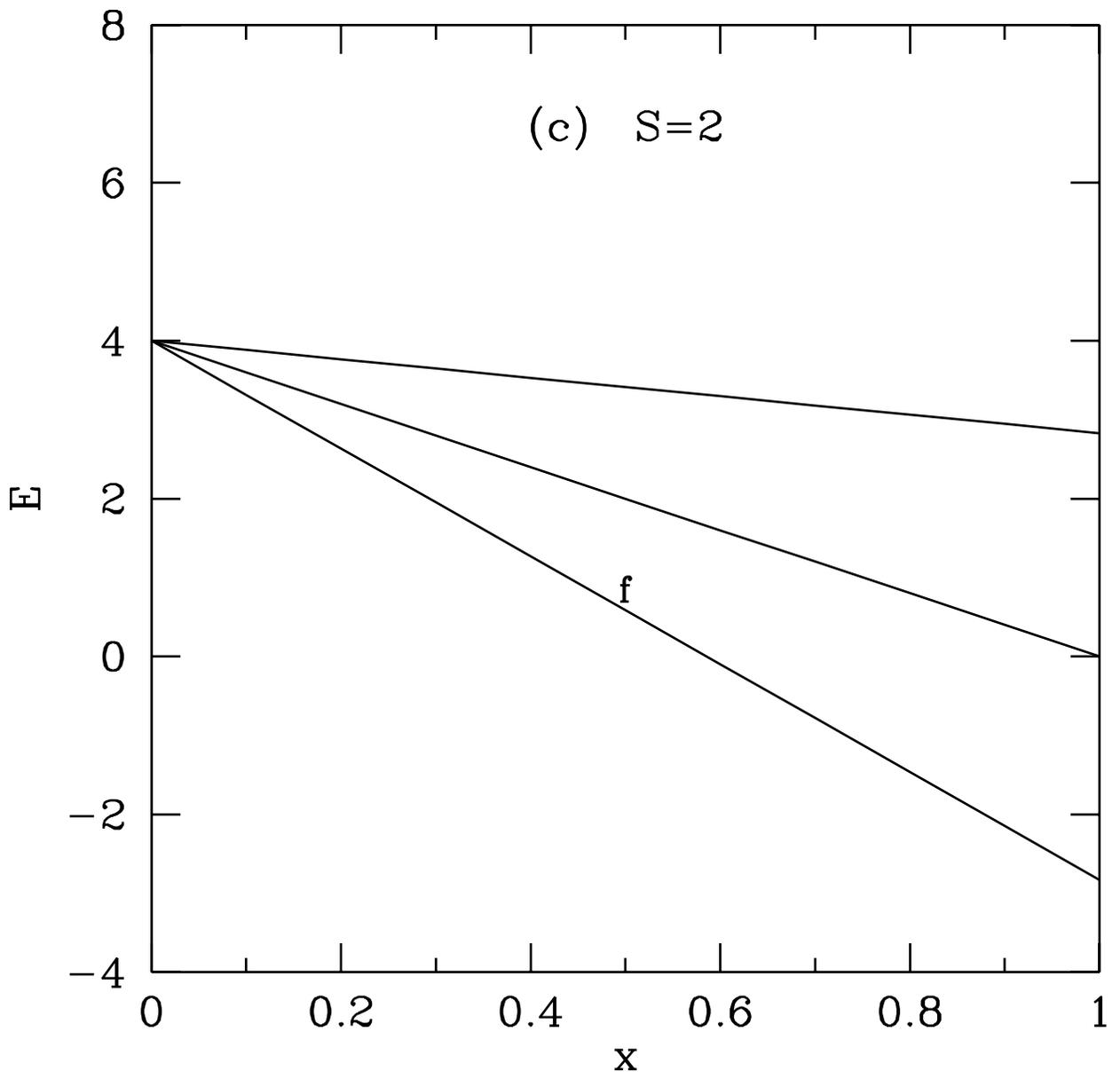
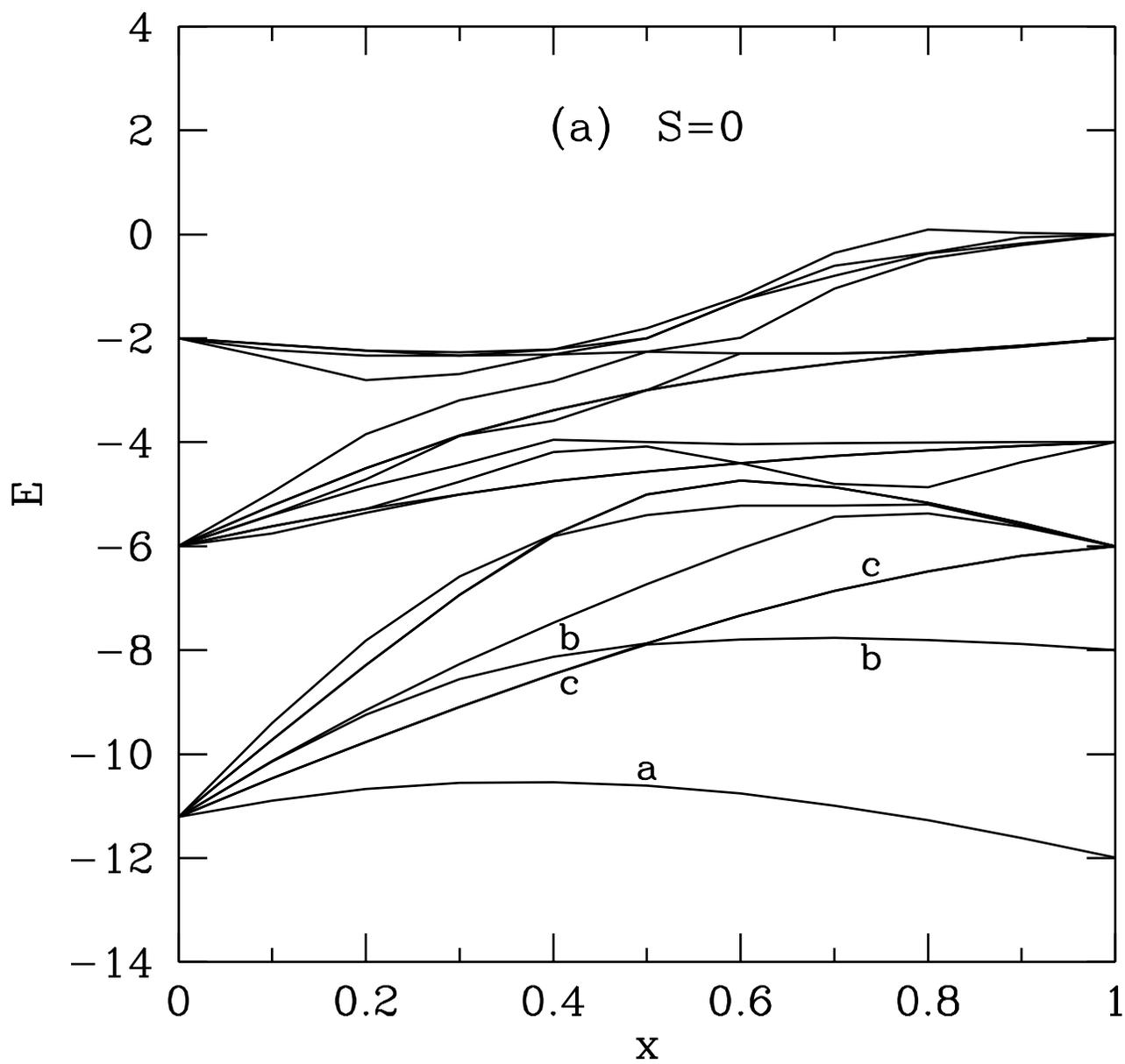
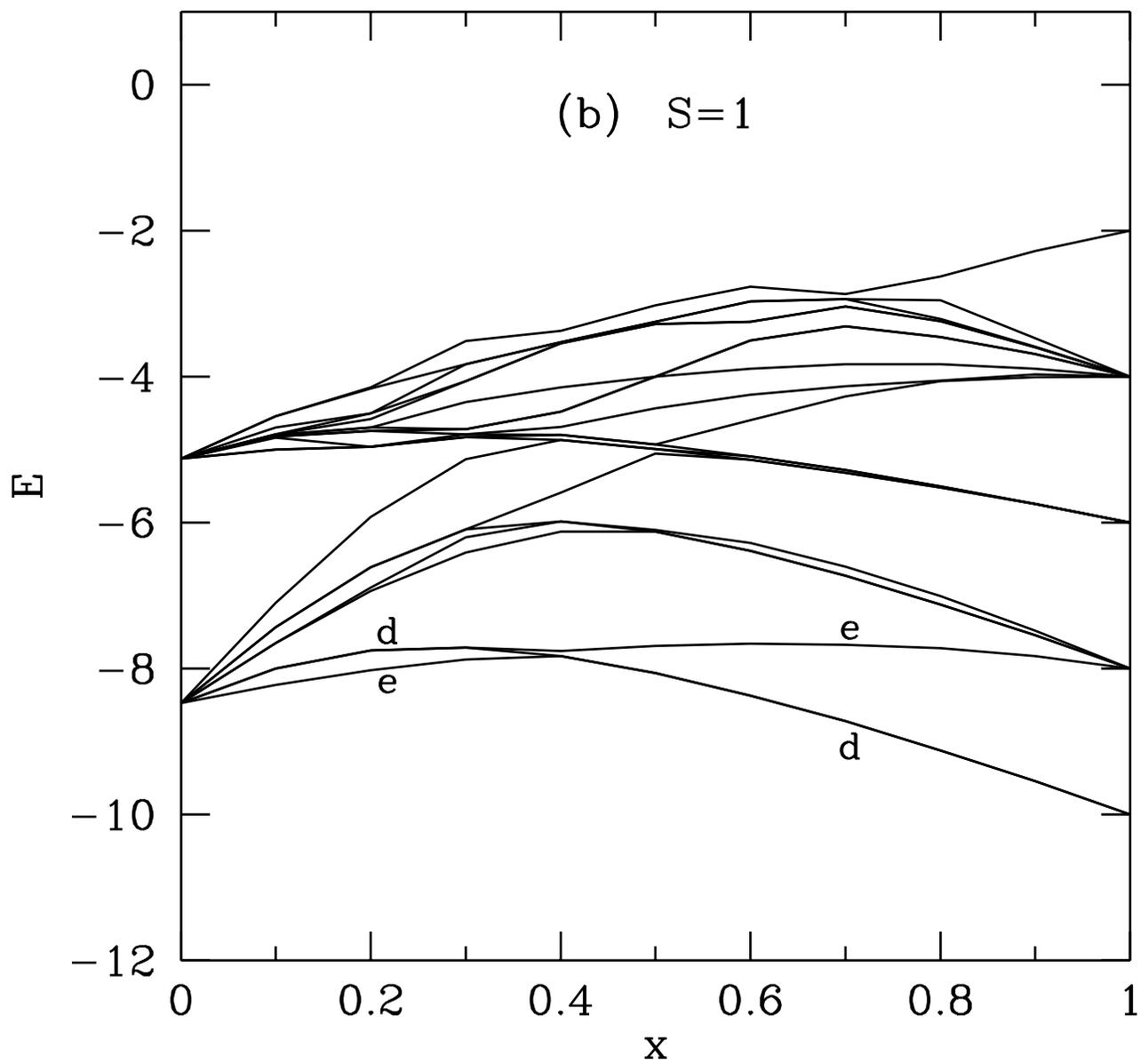
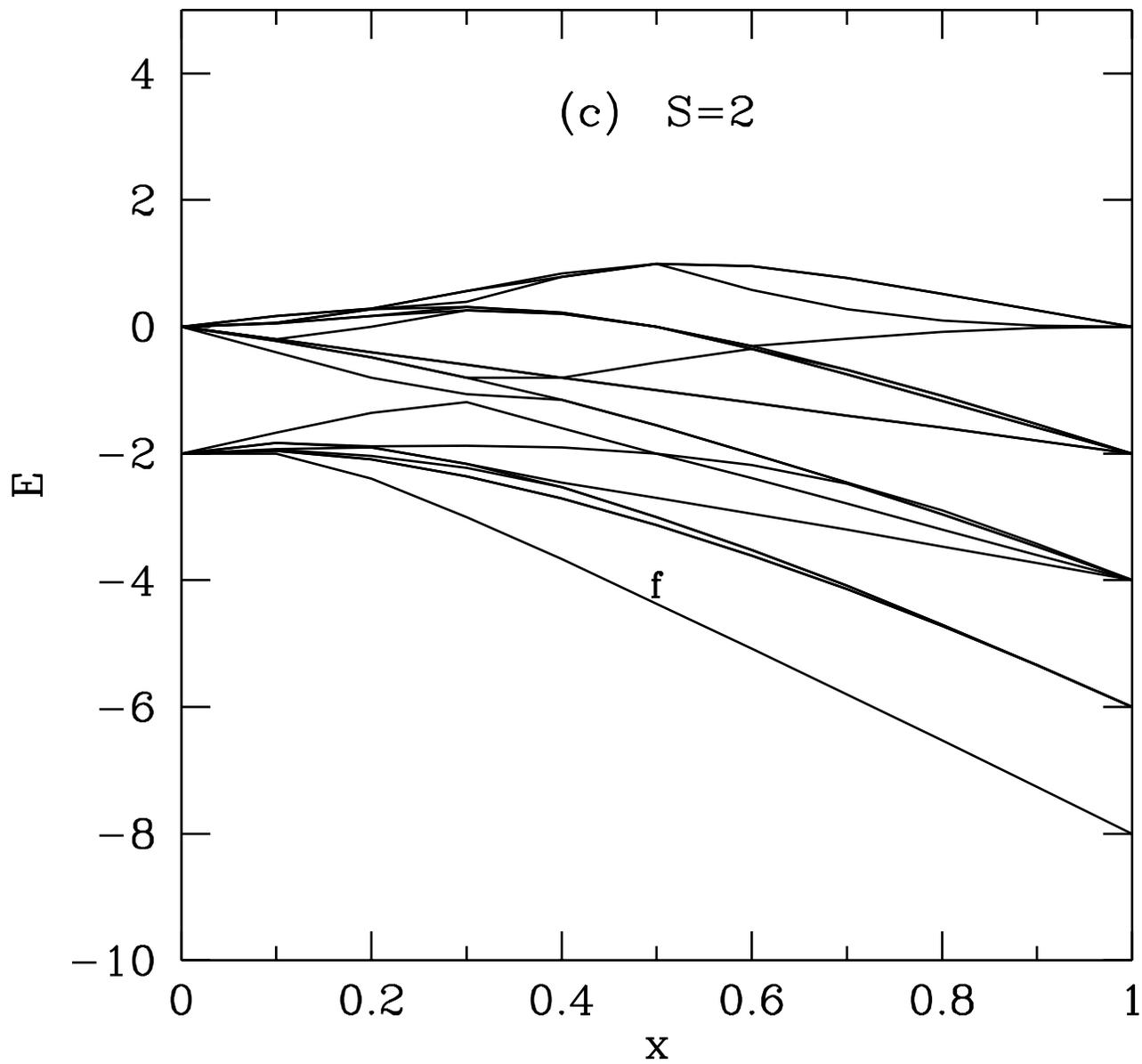
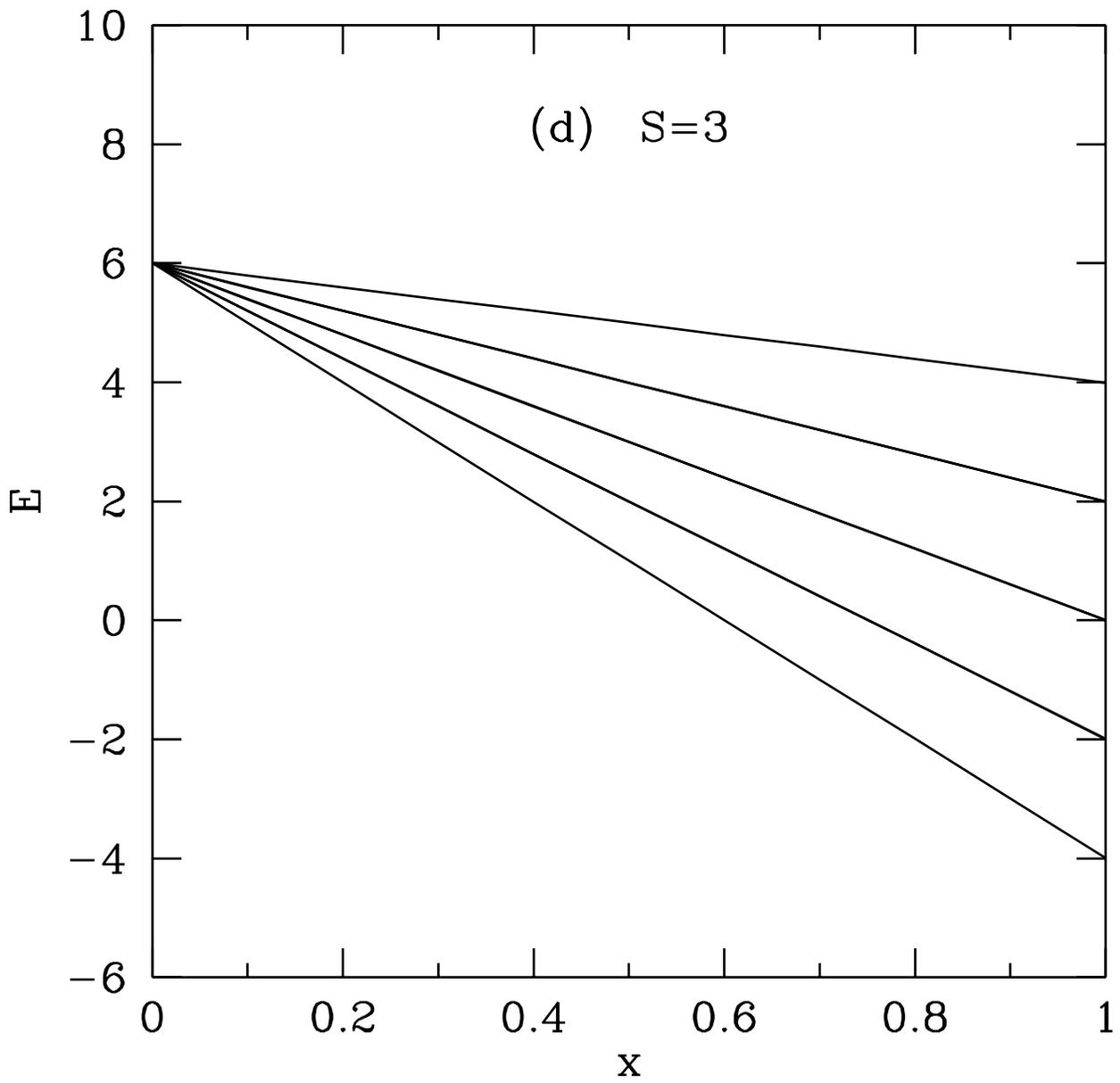
\end{document}